# Trion Hall effect in electron-hole double layers

Raghav Chaturvedi[1], Phuong X. Nguyen[1,2], Patrick Knüppel[1], Kenji Watanabe[3], Takashi Taniguchi[3], Kin Fai Mak[1,2,4,5*], Jie Shan[1,2,4,5*]

[1]School of Applied and Engineering Physics, Cornell University, Ithaca, NY, USA
[2]Kavli Institute at Cornell for Nanoscale Science, Ithaca, NY, USA
[3]National Institute for Materials Science, Tsukuba, Japan
[4]Laboratory of Atomic and Solid State Physics, Cornell University, Ithaca, NY, USA
[5]Max Planck Institute for the Structure and Dynamics of Matter, Hamburg, Germany.

[*]Email: kin-fai.mak@mpsd.mpg.de; jie.shan@mpsd.mpg.de

These authors contributed equally: Raghav Chaturvedi, Phuong X. Nguyen

**The realization of Coulomb-coupled electron-hole double layers in 2D semiconductor heterostructures has enabled the thermodynamic and transport studies of equilibrium exciton fluids without a magnetic field. By doping the exciton fluid with additional electrons/holes, an equilibrium fluid of trions – three-particle bound states of electrons and holes – further emerge, providing the platform to explore new transport phenomena associated with such composite particles. Here we report the observation of a Hall effect for trions in $MoSe_2/WSe_2$ heterobilayers, which support Coulomb-coupled electron and hole fluids with tunable densities. The Hall effect arises from a Lorentz force on trions under a perpendicular magnetic field. It is manifested in both Hall drag measurements and standard Hall effect measurements on just one of the semiconductor layers. For negatively charged trions, an electron-Hall effect is observed even in a hole-doped $WSe_2$ monolayer due to the presence of trion drags. The trion Hall effect also disappears when the trions are ionized at elevated temperatures and/or high trion densities. Our work opens the door for realizing quantum oscillations and the quantum Hall effect for trions.**

## Main

An equilibrium fluid of excitons – bound electron-hole pairs – is an ideal platform to explore bosonic quantum states of matter in solid-state materials (e.g. exciton condensation and superfluidity[1-14], and correlated insulating states of excitons[15-20]). Whereas earlier realizations have focused on the partially filled Landau levels in the quantum Hall regime in GaAs quantum wells[7,21,22] and graphene[23-25], recent works on Coulomb-coupled electron-hole double layers built on bilayer graphene[26] and 2D semiconductor heterostructures[27-33] have realized equilibrium excitons without a magnetic field. The latter setup consists of an electron-doped and a hole-doped 2D semiconductor layers separated by a thin layer of hexagonal boron nitride (hBN) that quenches interlayer tunneling (Fig. 1a). The application of an interlayer bias voltage ($V_b$) separates the

electron- and hole-chemical potentials and reduces the effective charge gap ($E_g - V_b$) of the heterobilayer from its zero-bias value $E_g$ (Fig. 1b). When $E_g - V_b$ is smaller than the exciton binding energy $E_b$, spontaneous emergence of an equilibrium fluid of interlayer (or dipolar) excitons has been demonstrated by thermodynamic and Coulomb drag studies[28-33].

The physics is further enriched by doping the exciton fluid with additional electrons/holes, forming a Bose-Fermi mixture[20,34,35]. Recent experiments have demonstrated the emergence of an equilibrium fluid of trions[36,37] when the electron density ($n$) in one layer is twice the hole density ($p$) in the other layer or vice versa (Fig. 1a,b). Since trions are either negatively (for $n = 2p$) or positively (for $p = 2n$) charged, they experience a Lorentz force under a perpendicular magnetic field ($B$) as shown in Fig. 1c; a Hall effect for these composite particles is therefore expected but has not been observed.

In this work, we first map out the phase diagram of the electron-hole double layer under a finite $B$-field and identify regions in the phase diagram that host a trion fluid. We then demonstrate a Hall effect for trions using both Hall drag measurements and standard Hall effect measurements on just the hole layer. Intriguingly, an electron-Hall effect is observed in the latter measurement due to the presence of a fluid of negative trions. We further show that the trion Hall effect disappears when the trions are ionized at elevated temperatures and/or high trion densities.

**Phase diagram under a finite *B*-field**

As shown in Fig. 1a, we study dual-gated electron-hole double layer devices made of monolayer $WSe_2$ (the hole layer or W-layer) and bilayer $MoSe_2$ (the electron layer or Mo-layer). (We used bilayer instead of monolayer $MoSe_2$ because of its better mechanical stability during the layer stacking process.) As detailed in earlier studies[28,30,36], we applied a constant perpendicular electric field with the gates to heavily dope the contact regions (with thicker hBN spacer) by electrons and holes to reduce the electrical contact resistance (Extended Data Fig. 1). Because the hBN thickness for the top and bottom gates is nearly the same, the antisymmetric combination $\Delta \equiv \frac{V_{bg}-V_{tg}}{2}$ of the top and bottom gate voltages ($V_{tg}$ and $V_{bg}$, respectively) is proportional to the $E$-field and the symmetric combination $V_g \equiv \frac{V_{bg}+V_{tg}}{2}$ tunes the net doping density of the double layer. The semiconductor layers are also separately contacted by metallic electrodes to access both the longitudinal ($R_{xx}$) and Hall ($R_{xy}$) resistances of each layer; the longitudinal ($R_{xx(drag)}$) and Hall ($R_{xy(drag)}$) drag resistances can also be measured. Both Coulomb drag measurements and measurements on a single layer are sensitive to interlayer Coulomb interactions and the formation of interlayer bound states[30-33,36,37]. (See Methods for details on device fabrication and electrical measurements.)

Figure 1d shows $R_{xx}$ of the W-layer as a function of $V_g$ and $V_b$ at zero $B$-field. The Mo-layer is kept in an open-circuit configuration (inset) so that no charge current is allowed to flow through. We focus on the W-layer for $R_{xx}$ and $R_{xy}$ measurements because of its better electrical contacts. Unless otherwise specified, the sample temperature is at $T = 1.5$K. Consistent with earlier studies[30-33,36,37], the W-layer is turned on in the $pi$ and $pn$ regions of the phase diagram. Here $p$, $i$ and $n$ denote hole-doping, intrinsic and electron-doping, respectively; the first and second entry denotes the doping state of the W- and Mo-layer, respectively. The dashed lines in Fig. 1d separate the different doping regions of the double layer. (The boundary separating $pi$ and $ii$ is a vertical line because the W-layer is electrically grounded.) An excitonic insulator (EI) region hosting a fluid of excitons with diverging $R_{xx}$ is marked by an enclosed triangle that protrudes into the $pn$ region. In addition, a narrow region hosting a fluid of trions with large $R_{xx}$ centered at the $p = 2n$ line is observed. In both the EI and trion regions, an insulating behavior for $R_{xx}$ (i.e. increasing $R_{xx}$ with decreasing $T$) is observed due to the formation of interlayer bound states and the open-circuit configuration in the Mo-layer. The results are fully consistent with Ref. [30-33,36,37].

Next, we examine the $R_{xx}$ map at $B = 4$T in Fig. 1e. Compared to the zero $B$-field map, the narrow region hosting trions centered at $p = 2n$ now expands to two triangular regions, ($X+X^+$) and ($X+X^-$), with large $R_{xx}$ on both sides of the EI region schematically illustrated in Fig. 1f. Quantum oscillations in $R_{xx}$ due to the formation of Landau levels are also observed. Correlated with the $R_{xx}$ map, large $R_{xx(drag)}$ and drag currents are also observed in both the EI region and the new triangular regions as shown in Fig. 2a and Extended Data Fig. 8. (As shown in the inset, we drive a current in the Mo-layer and measure the drag voltage in the W-layer to obtain $R_{xx(drag)}$.) The results suggest the formation of interlayer bound states in these regions. The interlayer bound states disappear at sufficiently high $V_b$ or $V_g$, where $R_{xx(drag)}$ becomes negligible and $R_{xx}$ reduces to that of an isolated $WSe_2$ monolayer, due to free carrier screening of the interlayer Coulomb interaction[30,36].

To understand the nature of the interlayer bound states in the ($X+X^+$)- and ($X+X^-$)-regions, we perform handedness-resolved optical reflection contrast (RC) studies near the fundamental exciton resonances in both the W- and Mo-layers at $B = 6$T (Methods). We first focus on a line cut at constant $V_g \approx -0.06$V in the phase diagram and plot the left- ($\sigma^+$) and right- ($\sigma^-$) handed RC spectrum as a function of $V_b$ to illustrate the main idea (Fig. 2b). The doping dependence of the attractive polaron resonance in each handedness channel gives information about the valley occupation (K or K') in the electron- and hole-layers[38-40].

With increasing $V_b$ in the $pi$ region, the RC spectrum is independent of $V_b$ because the Mo-layer, where $V_b$ is applied, is charge neutral and the W-layer is grounded; only the K'-valley of the W-layer is doped with holes under $B = 6$T (I in Fig. 2d), as revealed by the presence of both the repulsive and attractive polarons in only the $\sigma^-$ channel of the W-layer. As $V_b$ enters the $pn$ region, the repulsive polaron weakens quickly with $V_b$ in both layers and in both handedness channels; we will focus on the attractive polaron from now on. Just across the *pn* threshold near $V_b = 0.84$V (yellow dashed line), the attractive polaron appears in the $\sigma^+$ channel of the Mo-layer and is strengthened in the $\sigma^-$ channel of the W-layer; this corresponds to an onset of electron doping in the K-valley of the Mo-layer and an increase in hole doping in the K'-valley of the W-layer (II in Fig. 2d). The K-valley of the W-layer remains empty until $V_b$ reaches $V_b \approx 0.86$V (red dashed line), where the attractive polaron appears in the $\sigma^+$ channel, signaling the onset of hole doping in the K-valley (III in Fig. 2d). Finally, when $V_b$ exceeds $V_b \approx 0.88$V (black dashed line), the attractive polaron appears in the $\sigma^-$ channel of the Mo-layer, and electrons start to occupy the K'-valley (IV in Fig. 2d).

We summarize the spectrally integrated RC of the attractive polarons in the W-layer as a function of $V_g$ and $V_b$ for both $\sigma^+$ and $\sigma^-$ in Fig. 2c. Similar results for the Mo-layer are shown in Extended Data Fig. 3. Regions with finite RC in $\sigma^+$ and $\sigma^-$ correspond to hole doping in the K- and K'-valley, respectively. The results in Fig. 2c show that holes occupy both valleys in the (X+X$^+$)-region (see Fig. 1f), where large $R_{xx}$ and $R_{xx(drag)}$ are observed (Fig. 1e and 2a). Moreover, compared to the expected injection threshold without interlayer Coulomb interactions for the K-valley (red solid line in Fig. 2c), the experimental injection threshold is lowered in the (X+X$^+$)-region. These results support the formation of positive interlayer trions (together with neutral interlayer excitons) in this region. Specifically, positive trions are energetically favorable when two holes in the bound state are in a spin-singlet configuration (i.e. holes occupy both valleys due to spin-valley locking). To gain the trion binding energy, the system spontaneously lowers the injection threshold for the K-valley to form a fluid of spin-singlet trions. (See Methods for an estimate of the trion binding energy based on the injection threshold shift.) This remains the case until the valley Zeeman energy exceeds the trion binding energy at high enough $B$-fields, beyond which the system can no longer gain energy by forming trions and the (X+X$^+$)-region disappears (Extended Data Fig. 4 and 6). Similar behaviors are also observed for the negative trions in the (X+X$^-$)-region (Extended Data Fig. 3), where $R_{xx}$ is much higher compared to the (X+X$^+$)-region because holes in negative trions are much more tightly bound.

**Trion Hall effect**

With the presence of trions established in the (X+X$^+$)- and (X+X$^-$)-regions under a finite $B$-field, we now examine $R_{xy}$ (Fig. 3a) and $R_{xy(drag)}$ (Fig. 3b) as a function of $V_g$ and $V_b$ at $B = 4$T. The measurement configurations are illustrated in the insets. Correlated with $R_{xx(drag)}$ in Fig. 2a, a large (negligible) $R_{xy(drag)}$ is observed inside (outside) the trion regions of the phase diagram; $R_{xy(drag)}$ also changes sign between positive and negative trions. A similar behavior is observed for $R_{xy}$ except $R_{xy}$ outside the trion regions is nonzero and provides a measure of the Hall density ($n_H = \frac{B}{eR_{xy}}$) of the W-layer ($e > 0$ is the electron charge). We also examine, in Fig. 3c, $R_{xy}$ as a function of $V_g$ and $B$ at a fixed $V_b \approx 0.88$V (red solid line in Fig. 3a). For $B \lesssim 2$T, the trion regions have not expanded to $V_b \approx 0.88$V yet (Extended Data Fig. 6); the Hall response here measures only $n_H$ of the W-layer. The trion regions emerge between $B \approx 2$T and $B \approx 6$T, where a large $R_{xy}$ is observed, and disappear for $B \gtrsim 6$T because the valley Zeeman energy exceeds the trion binding energy and trions are no longer stabilized; the Hall response for $B \gtrsim 6$T again measures only $n_H$ of the W-layer. Note that $R_{xy}$ cannot be reliably measured near $V_g = 0$V due to the diverging $R_{xx}$ of the EI.

We further study the $T$-dependence of $R_{xy}$ and $R_{xy(drag)}$ in Fig. 4a and 4c, respectively. Here we keep a constant $B$-field at $B = 4$T (red solid line in Fig. 3a) and scan $V_g$ at varying temperatures. The trion regions with large $R_{xy}$ and $R_{xy(drag)}$ emerge at $T \lesssim 8$K. We plot the $T$-dependence of $R_{xy}$ and $R_{xy(drag)}$ at two selected line cuts in Fig. 4b and 4d, respectively (one for the positive and the other for the negative trion side as shown by the dashed lines in Fig. 4a and 4c). A negligible $R_{xy(drag)}$ is observed at $T \gtrsim 8$K, and below about 8K, $R_{xy(drag)}$ grows in magnitude and shows opposite signs for positive and negative trions. A similar dependence is observed for $R_{xy}$ except that $R_{xy}$ is nonzero at high temperatures due to the Hall response of the free holes in the W-layer. Remarkably, $R_{xy}$ turns negative for $T \lesssim 4$K on the negative trion side, i.e. an electron-Hall effect emerges in the hole layer.

In Fig. 3d, we compare the effective Hall density $n_H = \frac{B}{eR_{xy}}$ as a function of $V_g$ at $T = 1.5$K and 10K. The dashed lines show the total electron (orange) and hole (blue) densities in the Mo- and W-layer, respectively (i.e. the densities determined by the gate capacitances). Aside from a weak anomaly near $V_g = 0$V (due to remnant electron-hole interactions), $n_H$ at $T = 10$K follows the expected dependence for the hole density in the W-layer (blue dashed line). In contrast, $n_H$ at $T = 1.5$K starts to deviate from the blue dashed line as $V_g$ enters the positive trion region near $V_g = -0.07$V, decreases monotonically towards zero as $V_g$ approaches the charge neutrality point at $V_g = 0$V, changes sign for $V_g > 0$V, and finally jumps back to the blue dashed line near $V_g = 0.06$V as $V_g$ leaves the negative trion region.

The observed dependence in Fig. 3d can be understood in terms of a trion Hall effect. We describe the Coulomb-coupled electron-hole transport by a four-by-four conductivity tensor that includes conductivity tensors of a Bose-Fermi mixture consisting of free electrons, free holes, excitons and trions (Methods). Inverting the conductivity tensor yields $R_{xy} \approx \frac{B}{n_h e}$ (or $n_H \approx n_h$) in the positive trion region ($n_h$ is the free hole density), i.e. the Hall response of the W-layer is determined only by the free holes. This is expected as bound holes in excitons or trions do not contribute to $R_{xy}$ for an open-circuit Mo-layer. Therefore, the drop in $n_H$ below the blue dashed line in Fig. 3d as the system enters the positive trion region can be understood as a loss of free holes to the formation of trions and excitons. The free hole density vanishes at the charge neutrality point at $V_g = 0\text{V}$ as all holes in the W-layer are bound to form excitons.

In the negative trion region, we have $R_{\text{xy}} \approx -\frac{B}{e}\left(\frac{1+f}{n_{\text{e}}} + \frac{f^2\left(1-\frac{\sigma_{\text{X}0}}{\sigma_{\text{e}0}}\right)}{n_{\text{T}}}\right)$, where $n_e \geq 0$ and $n_T \geq 0$ are the free electron and trion densities, respectively, and the dimensionless factor $f$ is determined by the ratio of exciton to trion conductivity (Methods). Whereas the result is less straightforward to interpret than the positive trion region, the negative Hall resistance shows an electron-Hall effect in a hole layer, which unambiguously demonstrates the presence of a trion Hall effect and trion drags. The negative Hall resistance persists until the system leaves the negative trion region at high electron doping densities, where all bound states (excitons and trions) are ionized, and $R_{xy}$ changes sign and once again provides a measure for the total hole density in the W-layer.

In addition to ionization of the bound states by screening at high electron/hole doping densities ($|V_g| \gtrsim 0.06\text{V}$), trions can also be ionized by thermal excitations. This explains the gradual evolution of the low-temperature Hall response dominated by trions to the high-temperature response dominated by free holes observed in Fig. 4a and 4b. The crossover temperature scale provides an estimate for the trion binding energy, which decreases gradually from ≈ 8K (or 0.7meV) near charge neutrality to zero near the critical doping densities at $|V_g| \approx 0.06\text{V}$. (See Extended Data Fig. 5 and 7 for $T$-dependence of RC and $R_{xx}$, respectively, that also support this picture.)

**Conclusion and outlook**

In conclusion, we report the observation of a trion Hall effect in an electron-/hole-doped exciton fluid realized in Coulomb-coupled $MoSe_2/WSe_2$ heterobilayers. The most dramatic manifestation of the trion Hall effect is the observation of an electron-Hall effect in a hole layer when negative trions are stabilized. The trion Hall effect disappears when trions are ionized either at high electron/hole doping densities or at temperatures above the trion ionization temperature. Finally, we note the presence of quantum oscillations in $R_{xx}$ in the positive trion region of the phase diagram (the vertical stripes in Fig. 1e and Extended Data Fig. 8). Consistent with the low Hall density in this region (Fig. 3), the vertical, instead of diagonal, stripes demonstrate the survival of exciton binding upon hole-doping the exciton fluid at charge neutrality. The quantum oscillations here are not trion quantum oscillations because the cyclotron energy for free holes at $B = 4\text{T}$ (≈ 0.8meV) is already comparable to the trion binding energy; recurring transitions between bound trions and quantum Hall states for the free holes are expected under this condition. The situation is thus similar to the recurring transitions between excitonic insulators and electron-hole decoupled quantum Hall states reported recently in the same system at charge neutrality[32,33,41,42]. To realize true trion quantum oscillations, a low $B$-field keeping the cyclotron energy substantially smaller than the trion binding energy is required. In other words, higher trion mobility in cleaner samples is needed in future studies to realize trion quantum oscillations and the trion quantum Hall effect.

## Methods

### Device fabrication

We fabricated our devices using the layer-by-layer dry transfer method described in Ref. [43]. The individual layers were first mechanically exfoliated onto 285nm $SiO_2$/Si substrates and then screened for appropriate geometry and thickness using an optical microscope. Each layer was sequentially picked up at $70^0$C using a polymer-based stamp consisting of a polycarbonate layer on a polypropylene-carbonate-coated polydimethylsiloxane block. The finished stack was then released onto another $SiO_2$/Si substrate with prepatterned Pt electrodes, and the residual polymer was dissolved in chloroform and isopropanol. The current path in the $WSe_2$ layer was defined by etching the top gate using standard electron-beam lithography followed by reactive ion etching with oxygen plasma (Oxford Plasmalab80Plus). Lastly, Bi contacts to the $MoSe_2$ layer were fabricated using another step of electron-beam lithography followed by physical vapor deposition in a thermal evaporator.

Extended Data Fig. 1 shows optical images of devices 1 and 2. The black and red solid lines denote the $WSe_2$ monolayer and the $MoSe_2$ bilayer, respectively. The top and bottom gates are outlined in dashed black and red lines, respectively. The overlap area between the gates and the semiconductor layers determines the measurement channel (orange shaded region). The $WSe_2$ and $MoSe_2$ layers are separated by a thin hBN barrier (1.5-2nm) inside the channel and by a thick hBN spacer (~10nm) right outside the channel. The latter contact region serves as an exciton contact and facilitates exciton injection into the channel. To achieve Ohmic contacts at low temperatures, the $WSe_2$ and $MoSe_2$ layers are contacted with Pt and Bi electrodes[44,45], respectively. Furthermore, the gates are maintained at a high antisymmetric gating condition that heavily dopes the metal-semiconductor contact regions. A schematic cross-section of the device is shown in Extended Data Fig 1c.

### Electrical measurements

Electrical transport measurements were performed using standard lock-in techniques in a closed-cycle He4 cryostat (Oxford TeslatronPT) with a base temperature of 1.5K and magnetic field up to 12T. We performed two different types of measurements on our electron-hole bilayer devices (Extended Data Fig. 2); see Ref. [30,32,36] for details. In the first measurement, we performed standard Hall bar measurements of $R_{xx}$ and $R_{xy}$ of the W-layer while keeping the Mo-layer in open-circuit. In the second type (Coulomb drag measurements), we biased a current in the Mo-layer and measured both the longitudinal and transverse voltage drops in the W-layer to obtain $R_{xx(drag)}$ and $R_{xy(drag)}$. In both cases, we biased the device with an AC voltage of 1mV (RMS) at 7.33Hz, and measured the bias current and the longitudinal and transverse voltage drops simultaneously using Stanford Research lock-in amplifiers (SR860 and SR830). Note that all voltages were pre-amplified by Ithaco DL1201 preamplifiers with an input impedance of 100MΩ before sending the signals to the lock-in amplifiers. In the Coulomb drag measurements, the AC bias voltage was applied through a 1:1 transformer, which was connected to a 10kΩ potentiometer that distributed the AC voltage equally to the two terminals of the Mo-layer. The AC coupling between the Mo- and W-layers was minimized in this way. The DC bias voltage $V_b$ was applied to the middle of the potentiometer to keep the terminals of the Mo-layer at the same potential.

**Optical measurements**

Optical measurements were performed in a closed-cycle He4 cryostat (Attodry 2100, Attocube) with a base temperature of 1.6K and magnetic field up to 9T. A tungsten halogen lamp (Thorlabs) was used as a broadband white light source. Its output was coupled to a single mode fiber and focused onto the devices under normal incidence by a low-temperature microscope objective (0.8 numerical aperture). The incident intensity was kept below 50nW/μm$^{-2}$ to minimize sample heating. The reflected light was spectrally resolved by a spectrometer and detected by a liquid nitrogen cooled silicon charge-coupled device (Princeton Instruments). The reflectance contrast spectrum ($RC = \frac{I-I_0}{I_0}$) was obtained by comparing the reflected light intensity, $I$, from the sample to a featureless background, $I_0$, obtained at high electron/hole doping densities. For polarization resolved measurements, the left ($\sigma^+$) and right ($\sigma^-$) circularly polarized light was generated by a combination of a linear polarizer and an achromatic quarter-wave plate.

**Estimation of the trion binding energy**

We estimate the trion binding energy ($E_{T,b}$) based on the difference between the expected injection threshold without Coulomb interactions and the observed injection threshold in reflection contrast measurements. When trions form, the injection threshold for carriers in the minority valley is reduced by $E_{T,b}$. Figure 2c shows that the shift in threshold at $B = 6$T is about 25mV in $V_b$. By taking into account both the quantum and the geometrical capacitances of the double layer, we can convert the threshold shift to $E_{T,b} \approx \frac{C_Q^{-1}}{C_Q^{-1}+C_{2L}^{-1}}(25\text{meV}) \approx 1.2\text{meV}$. Here the quantum capacitance is given by $C_Q = e^2 \frac{g_s g_v m_\text{h}}{2\pi\hbar^2} \approx 30.12\mu\text{F cm}^{-2}$, where $g_s = 1$ and $g_v = 2$ are the spin and valley degeneracies, respectively, and $m_\text{h} \approx 0.45 m_0$ is the effective mass for holes in $WSe_2$ ($m_0$ is the free electron mass). The interlayer geometrical capacitance is given by $C_{2L} = \frac{\epsilon_r \epsilon_0}{d} \approx 1.55\mu\text{F cm}^{-2}$, where $\epsilon_r \approx 3.5$ is the out-of-plane dielectric constant of hBN, $\epsilon_0$ is the vacuum permittivity, and $d \approx 2$nm is the interlayer separation. The estimated $E_{T,b} \approx 1.2$meV is in good agreement with the temperature scale and the Zeeman energy scale for the ionization of trions (see main text).

**Coulomb-coupled electron-hole transport**

Following Ref. [46], we model the transport properties of the bilayer system with free carriers, trions and excitons as current carrying particles by a conductivity tensor. We denote the 2-by-2 conductivity tensors of the exciton, trion, free electron and free hole by $\sigma_{\rm X}$, $\sigma_{\rm T}$, $\sigma_{\rm e}$ and $\sigma_{\rm h}$, respectively. In the case of electron trions the current density can be expressed as:

$$\begin{pmatrix} J_e \\ J_h \end{pmatrix} = \begin{pmatrix} \sigma_{\rm X} + 2\sigma_T + \sigma_e & -\sigma_{\rm X} - 2\sigma_T \\ -\sigma_{\rm X} - \sigma_T & \sigma_{\rm X} + \sigma_T \end{pmatrix} \begin{pmatrix} E_e \\ E_h \end{pmatrix}. \tag{1}$$

The current density ($J_e$ and $J_h$) and electric field ($E_e$ and $E_h$) are 2D in-plane vectors in the electron and hole layers. The current density in the electron layer $J_e$ responds to the electric field in both the electron ($E_e$) and hole ($E_h$) layers: The former carries contributions from excitons, electron trions and free electrons; and only excitons and trions contribute to the latter due to Coulomb drags. The factor of 2 in front of $\sigma_T$ is due to the presence of two electrons in an electron trion. The current density in the hole layer $J_h$ can be understood in a similar fashion. Note that the drag conductivity $\sigma_{\rm D}$ that describes the frictional drag between free electrons and free holes is negligible at low temperatures compared to $\sigma_{\rm X}$ and $\sigma_{\rm T}$. We can also write down a similar expression for the case of hole trions:

$$\begin{pmatrix} J_e \\ J_h \end{pmatrix} = \begin{pmatrix} \sigma_{\rm X} + \sigma_T & -\sigma_{\rm X} - \sigma_T \\ -\sigma_{\rm X} - 2\sigma_T & \sigma_{\rm X} + 2\sigma_T + \sigma_h \end{pmatrix} \begin{pmatrix} E_e \\ E_h \end{pmatrix}. \tag{2}$$

The 2-by-2 conductivity tensors under a perpendicular magnetic field are:

$$\sigma_{\rm X} = \frac{n_{\rm X} e^2 \tau_X}{m_X} \begin{pmatrix} 1 & 0 \\ 0 & 1 \end{pmatrix},\ \sigma_{\rm T} = \frac{n_{\rm T} e^2 \tau_T}{m_T} \begin{pmatrix} 1 & -\omega_{c,T}\tau_T \\ \omega_{c,T}\tau_T & 1 \end{pmatrix},$$
$$\sigma_{\rm e} = \frac{n_{\rm e} e^2 \tau_e}{m_e} \begin{pmatrix} 1 & -\omega_{c,e}\tau_e \\ \omega_{c,e}\tau_e & 1 \end{pmatrix},\ \sigma_{\rm h} = \frac{n_{\rm h} e^2 \tau_h}{m_h} \begin{pmatrix} 1 & -\omega_{c,h}\tau_h \\ \omega_{c,h}\tau_h & 1 \end{pmatrix}. \tag{3}$$

Here $n_{\rm X}$, $n_{\rm T}$, $n_{\rm e}$ and $n_{\rm h}$ are the densities for excitons, trions, free electrons and free holes, respectively; $\tau_{\rm X}$, $\tau_{\rm T}$, $\tau_{\rm e}$ and $\tau_{\rm h}$ are the transport scattering times; $m_{\rm X}$, $m_{\rm T}$, $m_{\rm e}$ and $m_{\rm h}$ are the particle masses; $\omega_{c,T}$, $\omega_{c,e}$ and $\omega_{c,h}$ are the cyclotron frequencies of trions, free electrons and free holes, respectively. To further simply the expressions, we note that the free hole (electron) conductivity is negligible in the case of electron (hole) trions because of the negligible $n_{\rm h}$ ($n_{\rm e}$). Furthermore, when standard Hall effect measurements are performed on the W-layer with the Mo-layer in open-circuit, Eqn. (1) is reduced to:

$$\begin{pmatrix} J_{ex} \\ J_{ey} \\ J_{hx} \\ J_{hy} \end{pmatrix} = \begin{pmatrix} 0 \\ 0 \\ J_{hx} \\ 0 \end{pmatrix} = \Sigma \begin{pmatrix} E_{ex} \\ E_{ey} \\ E_{hx} \\ E_{hy} \end{pmatrix} \Longrightarrow \begin{pmatrix} 0 \\ 0 \\ 1 \\ 0 \end{pmatrix} = \Sigma \begin{pmatrix} R_{xx(drag)} \\ R_{xy(drag)} \\ R_{xx} \\ R_{xy} \end{pmatrix}, \tag{4}$$

where $x$ and $y$ denote the longitudinal and transverse directions, respectively, and $\Sigma$ is a short-hand notation for the 4-by-4 conductivity tensor in Eqn. (1) and (2). Similarly, when Coulomb drag measurements are performed with a longitudinal driving current in the Mo-layer, we have

$$\begin{pmatrix} J_{ex} \\ J_{ey} \\ J_{hx} \\ J_{hy} \end{pmatrix} = \begin{pmatrix} J_{ex} \\ 0 \\ 0 \\ 0 \end{pmatrix} = \Sigma \begin{pmatrix} E_{ex} \\ E_{ex} \\ E_{hx} \\ E_{hy} \end{pmatrix} \Rightarrow \begin{pmatrix} 1 \\ 0 \\ 0 \\ 0 \end{pmatrix} = \Sigma \begin{pmatrix} R_{xx} \\ R_{xy} \\ R_{xx(drag)} \\ R_{xy(drag)} \end{pmatrix}. \tag{5}$$

Solving Eqn. (4) and Eq (5) gives the longitudinal and Hall resistances measured in our experiment. In particular, we have $R_{xy} \approx \frac{B}{n_h e}$ for hole trions and $R_{xy} \approx -\frac{B}{e}\left(\frac{1+f}{n_\mathrm{e}} + \frac{f^2\left(1-\frac{\sigma_{\mathrm{X}0}}{\sigma_{\mathrm{e}0}}\right)}{n_\mathrm{T}}\right)$ for electron trions with $f = \sigma_{T_0}/(\sigma_{T_0} + \sigma_{X_0})$, where $\sigma_{T_0} = \frac{n_\mathrm{T} e^2 \tau_T}{m_T}$ and $\sigma_{X_0} = \frac{n_\mathrm{X} e^2 \tau_X}{m_X}$.

## Figures

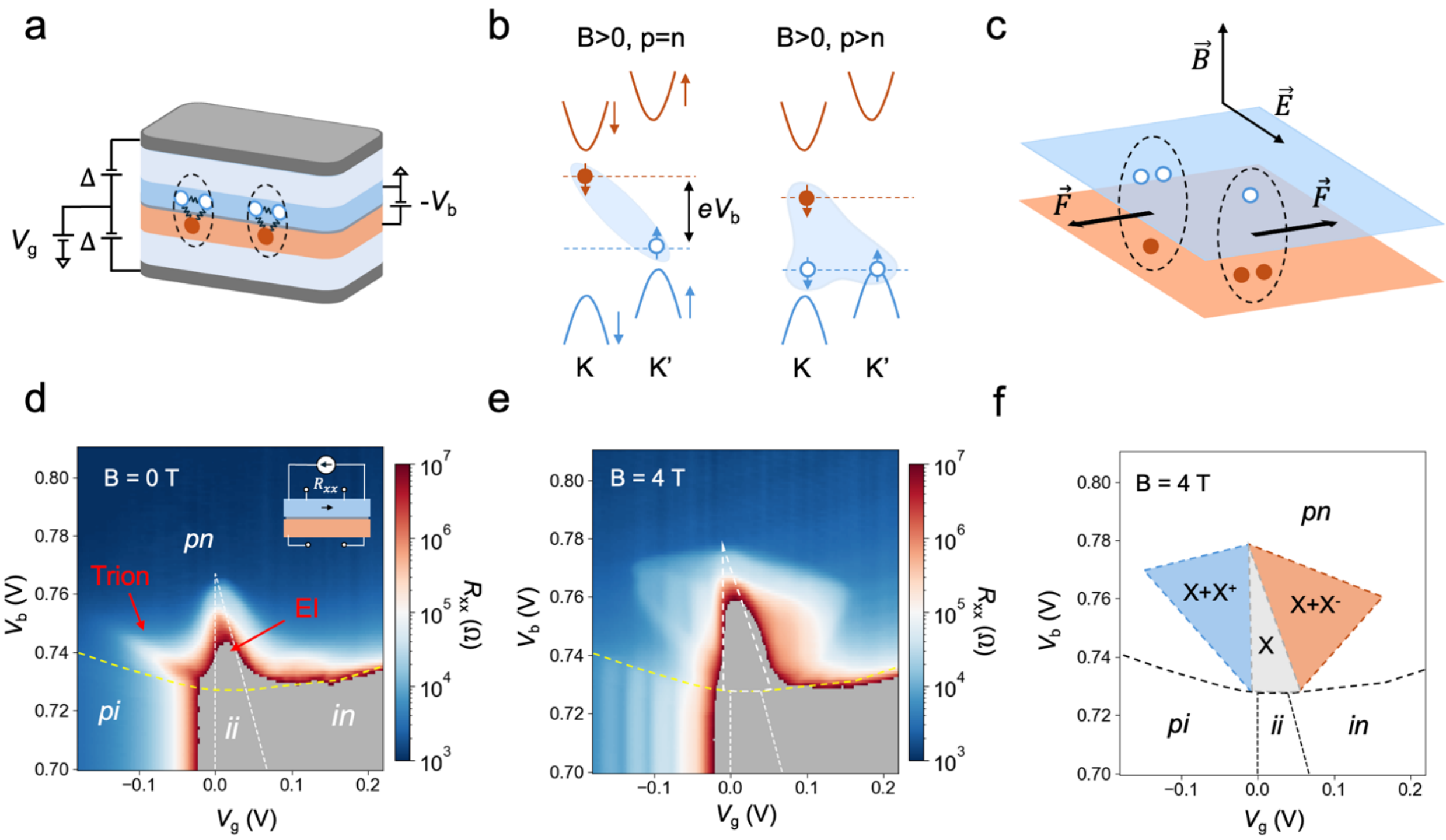

**Figure 1 | Trions in Coulomb-coupled electron-hole double layers. a,** Schematic dual-gated device structure supporting trions (three-particle bound states encircled by dashed lines). The electron (orange) and hole (blue) layers are separated by a thin (≈ 2nm) hBN barrier in the channel. $V_g$ and Δ denote the symmetric and antisymmetric combination of the gate voltages, respectively; $V_b$ is the bias voltage applied to the Mo-layer. **b,** Schematic type-II band alignment under a finite magnetic field with spin- and valley-split Mo-conduction bands (orange) and W-valence bands (blue). The electron- and hole-Fermi levels are split by $eV_b$. Excitons form when their binding energy exceeds the interlayer charge gap (left panel). Spin-singlet trions emerge in a doped excitonic insulator (right panel). **c,** Illustration of the trion Hall effect. Electron and hole trions experience opposite Lorentz forces under a perpendicular magnetic field and a bias electric field. **d,e,** $R_{xx}$ at $T = 1.5$K as a function of $V_g$ and $V_b$ under $B = 0$T (**d**) and 4T (**e**) (device 1). Schematic measurement configuration is shown in the inset of **d**. Dashed lines mark the phase boundaries separating the *pn*, *pi*, *ii*, and *in* regions (see main text). Red arrows in **d** mark the insulating regions with large $R_{xx}$. Grey-shaded areas denote completely insulating regions. **f,** Constructed electrostatics phase diagram under $B = 4$T. The EI, positive trion and negative trion regions with large $R_{xx}$ are denoted by X, X+X$^+$ and X+X$^-$, respectively.

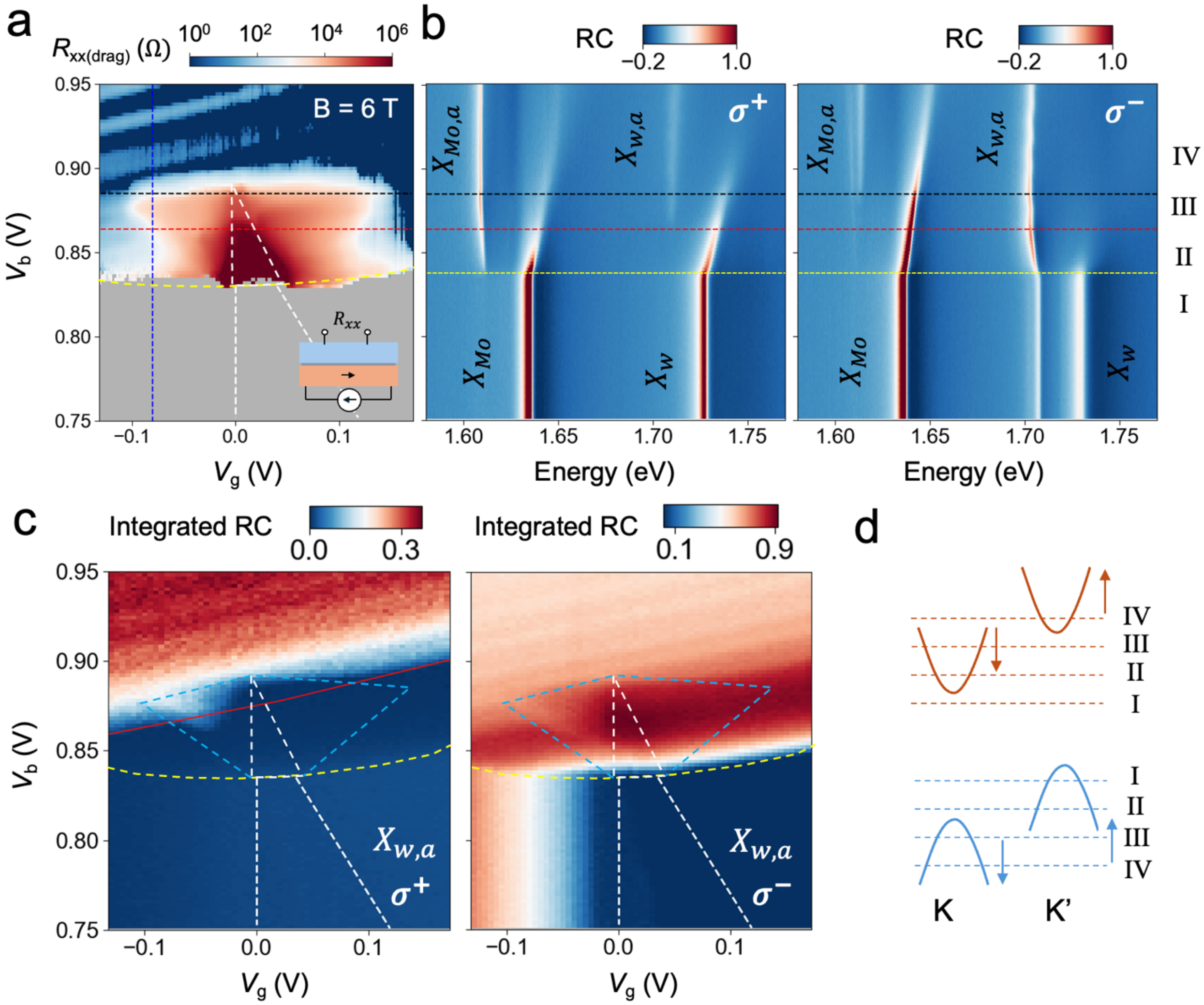

**Figure 2 | Phase diagram under finite magnetic fields. a,** $R_{xx(drag)}$ as a function of $V_g$ and $V_b$ at $B = 6$T and $T = 1.5$K (device 3). Schematic measurement configuration is shown in the inset. **b,** Reflection contrast (RC) spectra in the $\sigma_+$ (left) and $\sigma_-$ (right) channels measured along the blue dashed line in **a**. $X_{W(Mo)}$ and $X_{W(Mo),a}$ denote the neutral exciton (or repulsive polaron) and the attractive polaron resonances in the W(Mo)-layer, respectively. Red and black dashed lines mark the onset of occupation of both valleys in the W- and Mo-layers, respectively; yellow dashed line marks the onset of the *pn* region. Equilibrium positive trions (with large $R_{xx(drag)}$) emerge only in between the red and black dashed lines, i.e. when both valleys are occupied in the W-layer and only one valley is occupied in the Mo-layer. **c,** Integrated RC of the $X_{W,a}$ resonance in both the $\sigma_+$ (left) and $\sigma_-$ (right) channels as a function of $V_g$ and $V_b$. Dashed lines mark the phase boundaries corresponding to those in Fig. **1f**. The red solid line in the $\sigma_+$ channel marks the onset of doping in the absence of interlayer Coulomb interactions. **d,** Schematic type-II band alignment showing four different configurations (I, II, III and IV) of the electron and hole Fermi levels corresponding to the four regions marked in **b**.

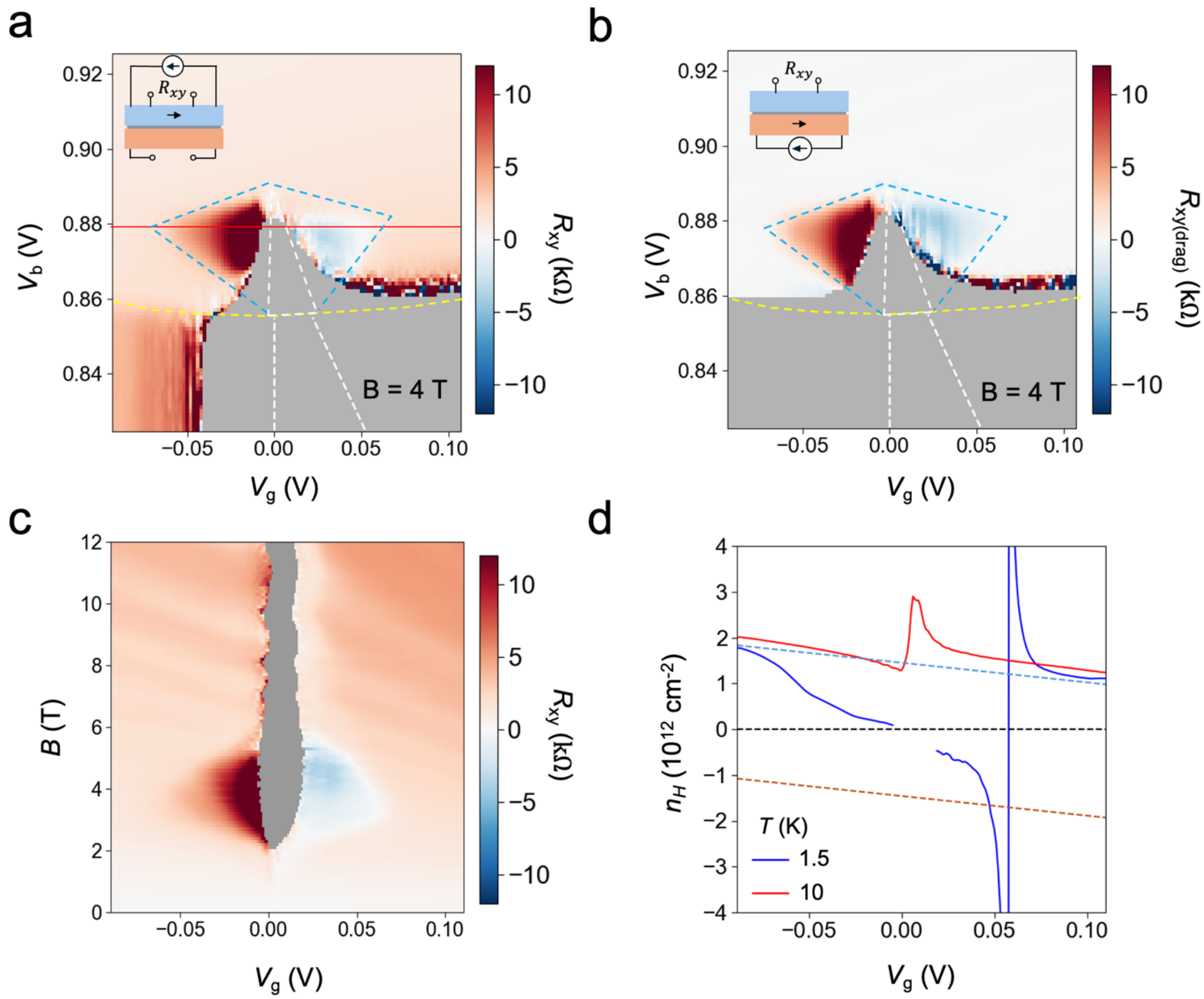

**Figure 3 | Trion Hall effect. a,b,** $R_{xy}$ (**a**) and $R_{xy(drag)}$ (**b**) as a function of $V_g$ and $V_b$ at $B = 4$T and $T = 1.5$K (device 2). Schematic measurement configurations are shown in the inset. Dashed lines mark the phase boundaries corresponding to those in Fig. **1f**. **c,** Magnetic field dependence of $R_{xy}$ along the red line in **a**. **d,** Hall density, $n_H$, plotted along the red line in **a** at $T = 1.5$K and 10K. The orange and blue dashed lines denote the expected electron and hole densities in the Mo- and W-layers, respectively, based on the calibrated gate capacitances.

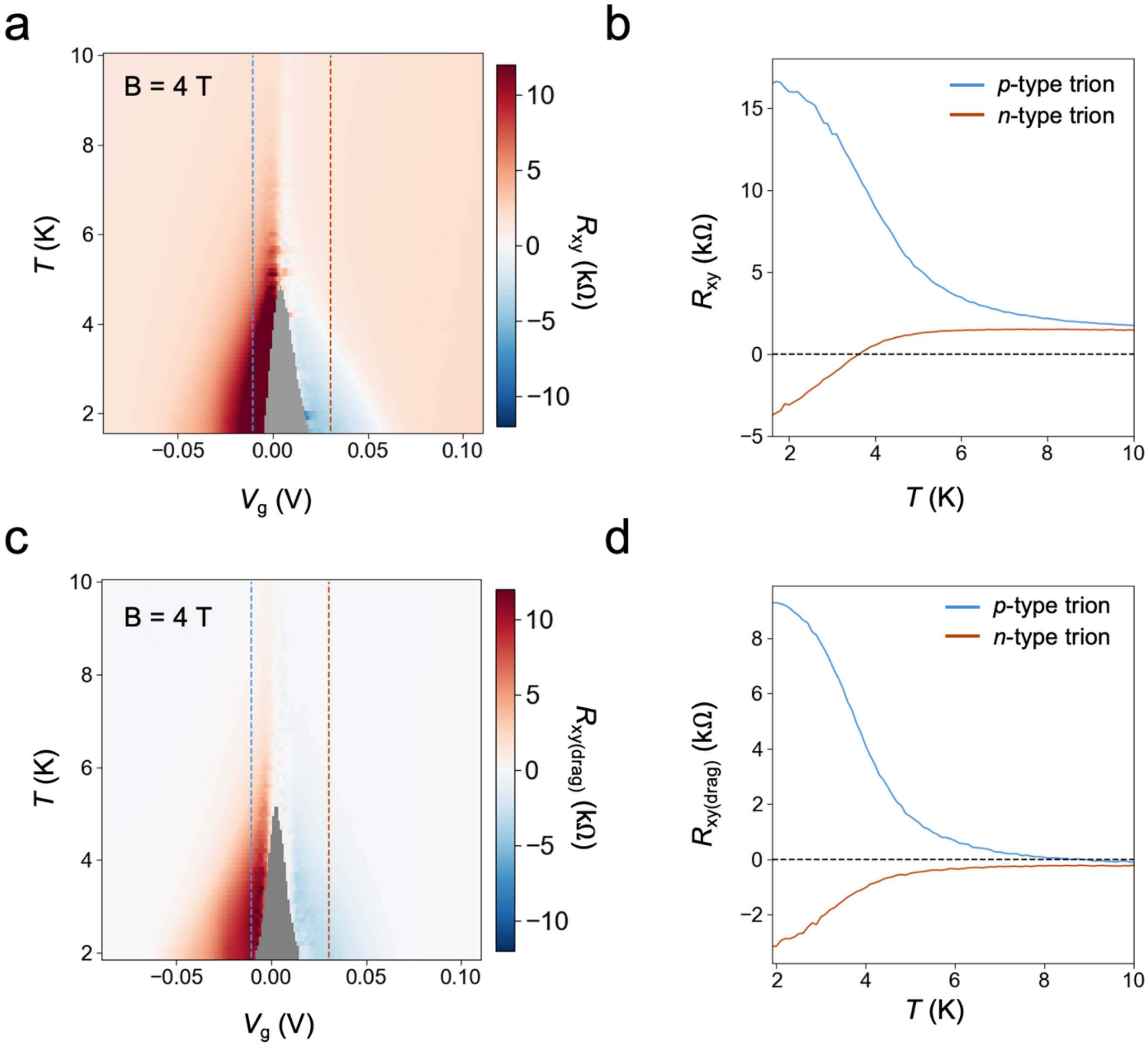

**Figure 4 | Temperature dependence of the trion Hall effect. a,c,** $R_{xy}$ (**a**) and $R_{xy(drag)}$ (**c**) as a function of $V_g$ and $T$ at $B = 4\text{T}$ along the red line in Fig. **3a**. **b,d,** Line cuts of **a** and **c** on the positive trion (blue line) and negative trion (orange line) side. A sign change in both $R_{xy}$ and $R_{xy(drag)}$ around charge neutrality (i.e. $V_g \approx 0\text{V}$) is observed for $T \lesssim 8\text{K}$ when trions emerge; $R_{xy(drag)}$ approaches zero at high temperatures as trions are ionized into free carriers.

## Extended Data Figures

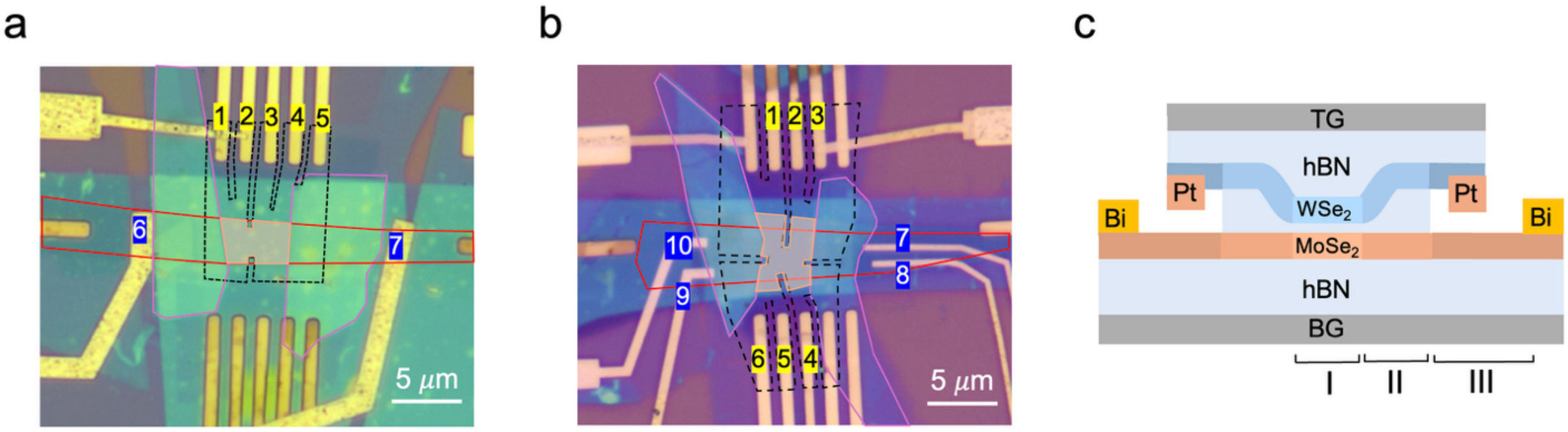

**Extended Data Figure 1 | Optical images and device schematic. a,b,** Optical micrographs of device 1 **(a)** and device 2 **(b)**. Top gate (black dashed line), bottom gate (red line) and exciton contact hBN (pink line) are outlined. The orange-shaded region marks the exciton channel. The Pt contacts to the W-layer and Bi contacts to the Mo-layer are numbered in yellow and blue, respectively. **c** Cross-section of the device showing doping densities across the device. The metal-semiconductor contact region, exciton contact region and channel region are labelled by III, II, and I, respectively. Lighter to darker shades imply increasing carrier density (blue for holes and orange for electrons). TG and BG denote the top and bottom gate, respectively.

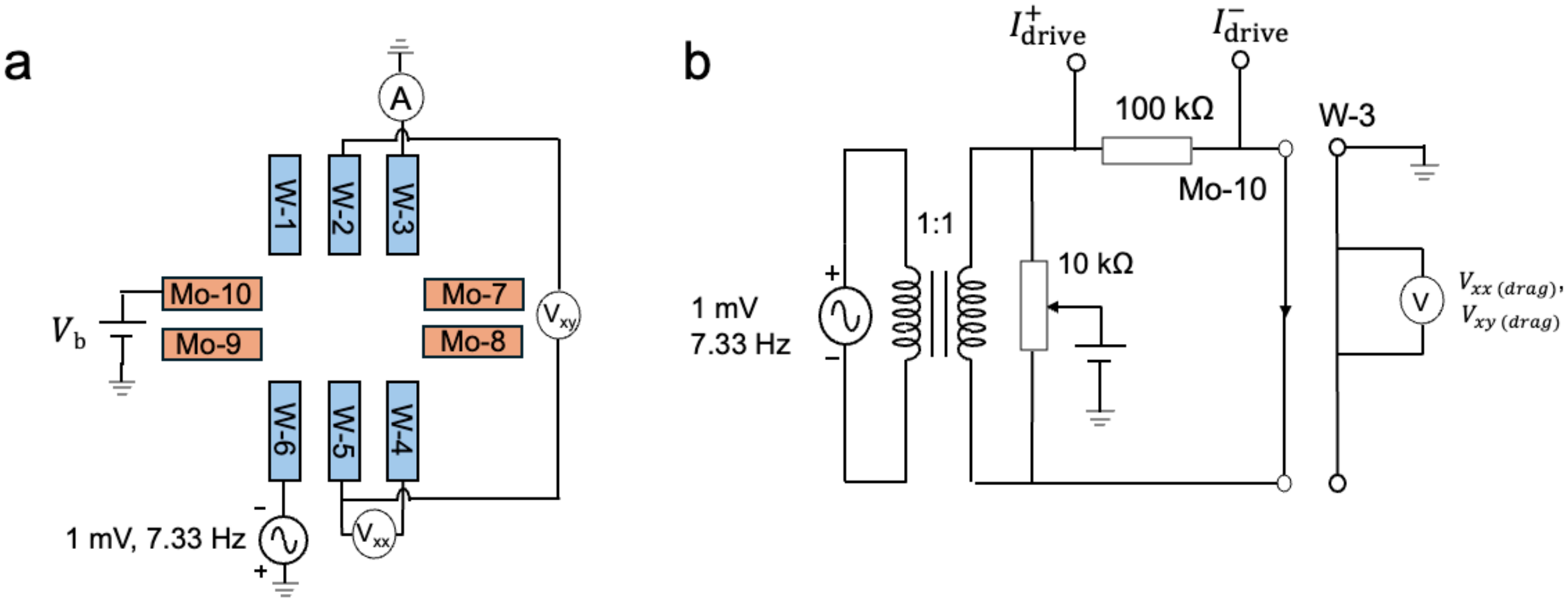

**Extended Data Figure 2 | Schematic measurement circuit. a,** Circuit diagram for open-circuit measurements of $R_{xx}$ and $R_{xy}$ in the W-layer. The Mo-layer is in open-circuit with only DC bias $V_b$ applied to one of the pins. $R_{xx}$ and $R_{xy}$ is measured using the connected pins as shown. **b,** Circuit diagram for measurements of $R_{xy(drag)}$ when an AC current is driven in the Mo-layer via a 1:1 transformer. The DC bias $V_b$ is applied using a potentiometer tuned to minimize the interlayer AC coupling. The Mo-layer is connected to a 100kΩ resistor in series for measurements of the drive current. $R_{xx(drag)}$ and $R_{xy(drag)}$ are then measured with the W-layer in open-circuit.

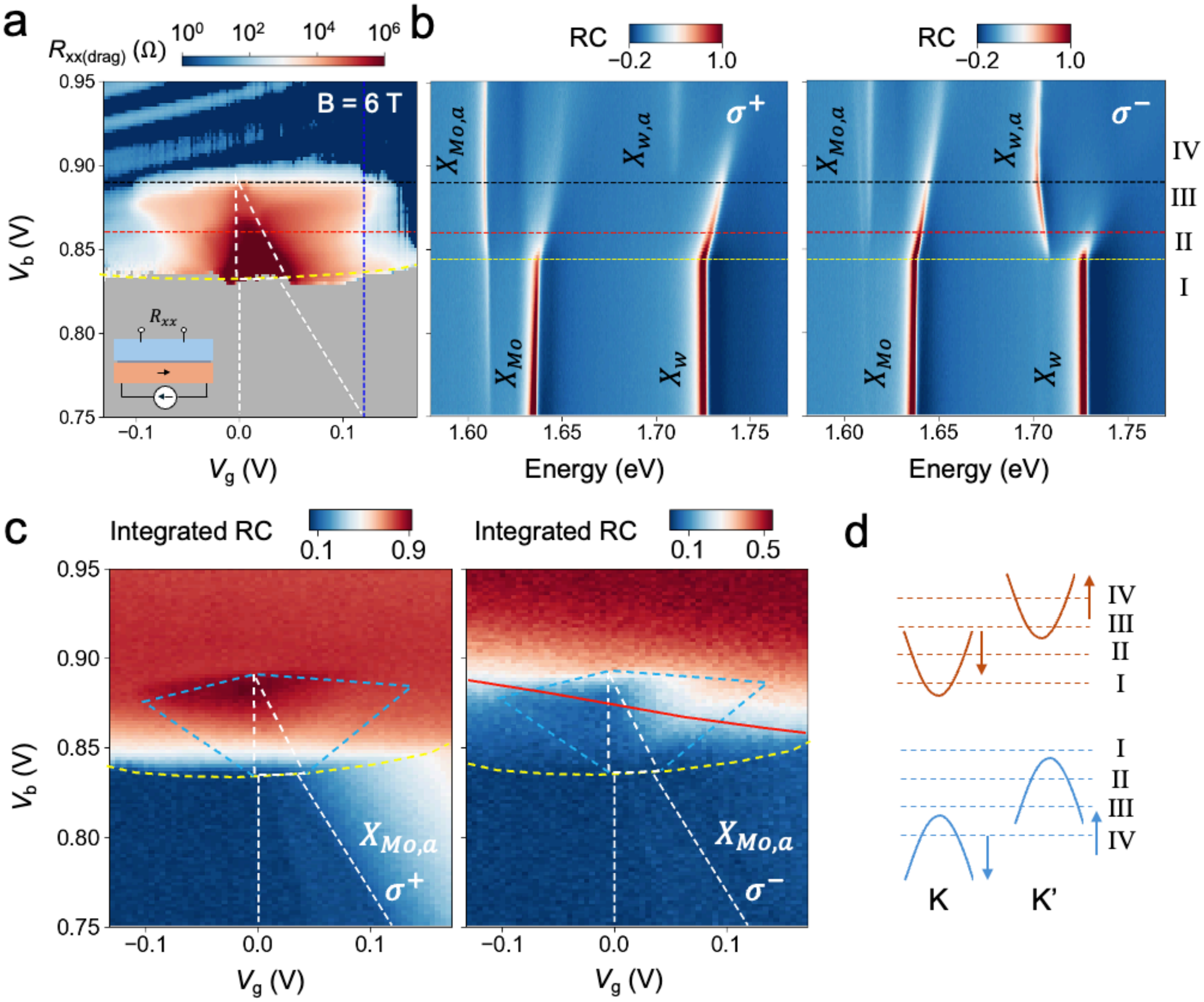

**Extended Data Figure 3 | Analysis of $X_{Mo,a}$ corresponding to Figure 2. a,** $R_{xx(drag)}$ as a function of $V_g$ and $V_b$ at $B = 6$T and $T = 1.5$K (device 3). **b,** RC spectra in the $\sigma_+$ (left) and $\sigma_-$ (right) channels measured along the blue dashed line in **a**. Red and black dashed lines mark the onset of occupation of both valleys in the Mo- and W-layers, respectively; yellow dashed line marks the onset of the *pn* region. Equilibrium negative trions (with large $R_{xx(drag)}$) emerge only in between the red and black dashed lines, i.e. when both valleys are occupied in the Mo-layer and only one valley is occupied in the W-layer. **c,** Integrated RC of the $X_{Mo,a}$ resonance in both the $\sigma_+$ (left) and $\sigma_-$ (right) channels as a function of $V_g$ and $V_b$. The red solid line in the $\sigma_-$ channel marks the onset of doping in the absence of interlayer Coulomb interactions. **d,** Schematic type-II band alignment showing four different configurations (I, II, III and IV) of the electron and hole Fermi levels corresponding to the four regions marked in **b**.

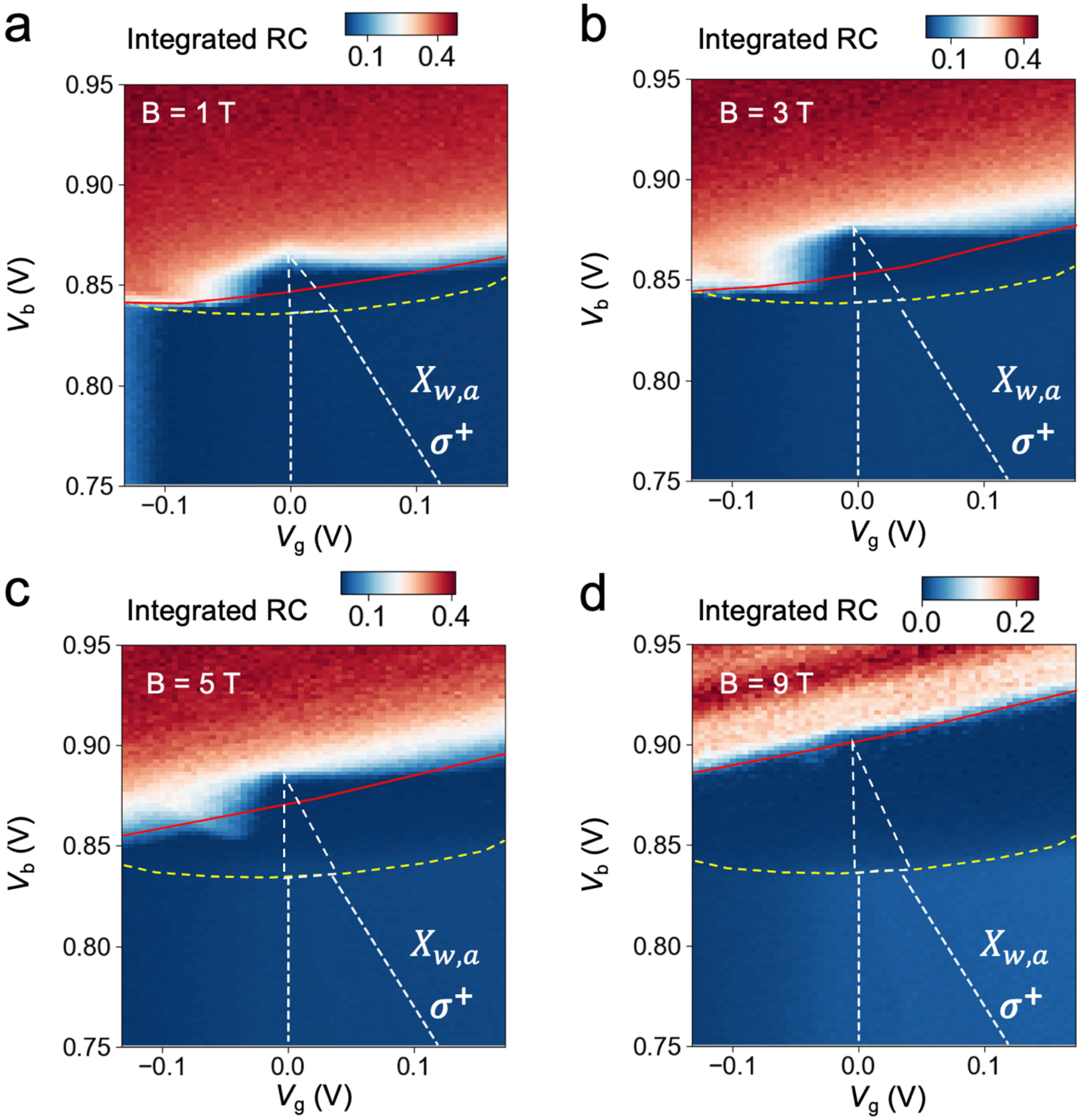

**Extended Data Figure 4 | Reflection contrast of $X_{W,a}$ under varying magnetic fields. a-d,** Integrated RC of the $X_{W,a}$ resonance in the $\sigma_+$ channel as a function of $V_g$ and $V_b$ at $T = 1.6$K and $B = 1$T (**a**), 3T (**b**), 5T (**c**) and 9T (**d**). Dashed lines mark the phase boundaries corresponding to those in Fig. **1f**. Red solid lines mark the onset of doping in the absence of interlayer Coulomb interactions. The regions hosting spin-singlet trions disappear under high fields due to the large spin-valley Zeeman effect that polarizes the electron/hole spins.

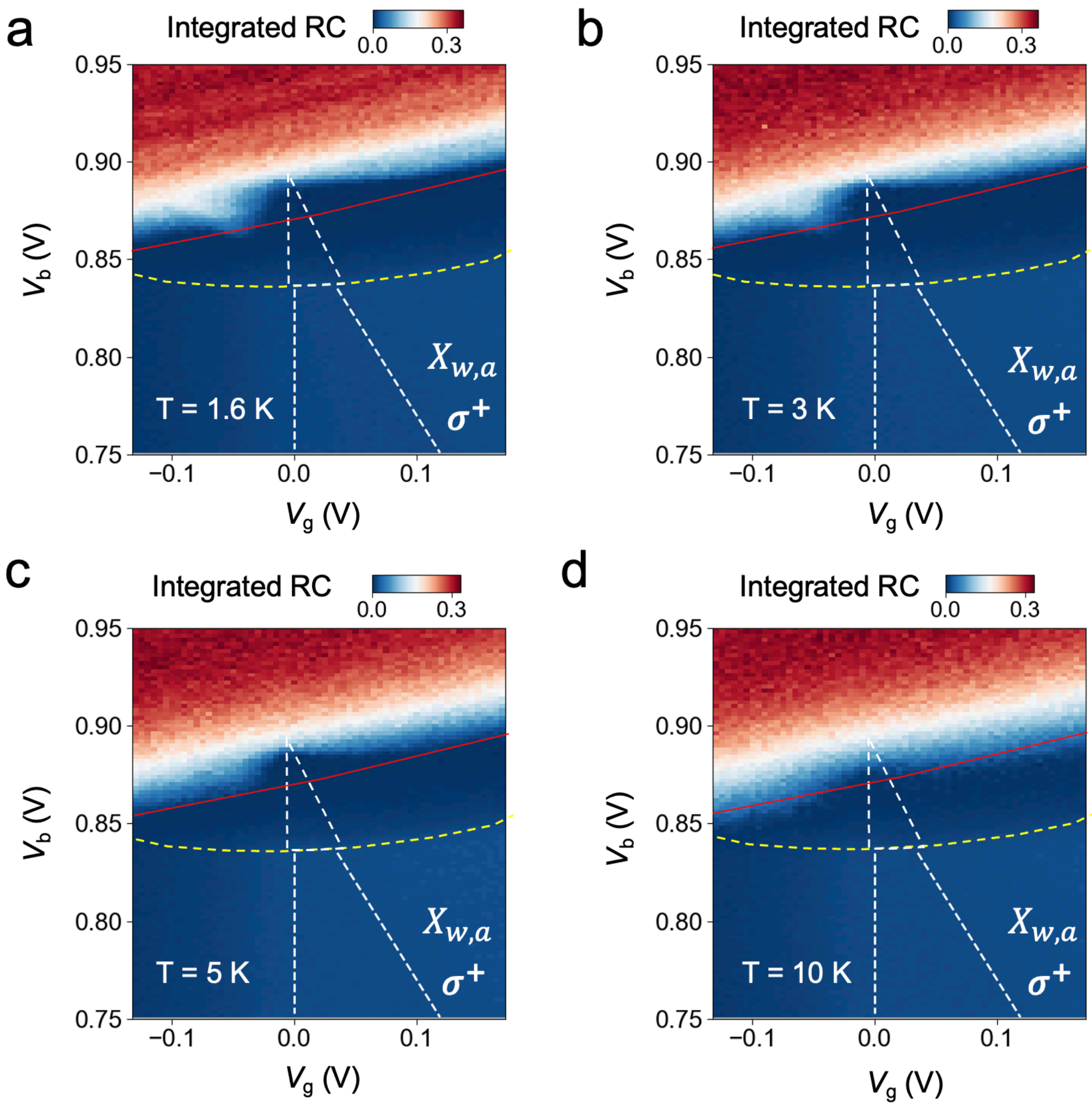

**Extended Data Figure 5 | Reflection contrast of $X_{W,a}$ under varying temperatures. a-d,** Integrated RC of the $X_{W,a}$ resonance in the $\sigma_+$ channel as a function of $V_g$ and $V_b$ at $B = 6$T and $T = 1.6$K (**a**), 3K (**b**), 5K (**c**) and 10K (**d**). Dashed lines mark the phase boundaries corresponding to those in Fig. **1f**. Red solid lines mark the onset of doping in the absence of interlayer Coulomb interactions. At 10K, trions are ionized and the doping threshold nearly follows the red solid line.

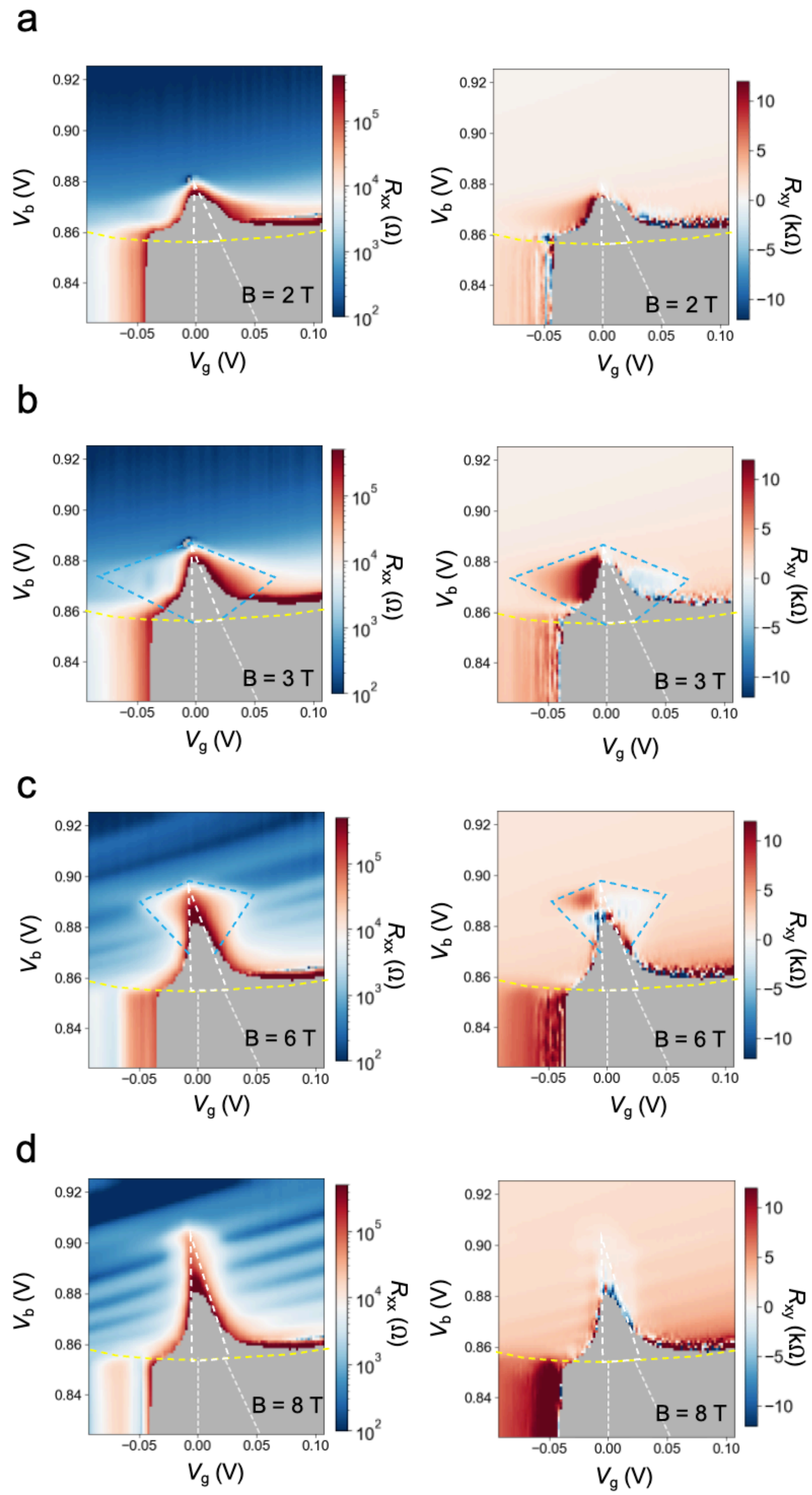

**Extended Data Figure 6 | Magnetic field evolution of $R_{xx}$ and $R_{xy}$. a-d,** $R_{xx}$ (left) and $R_{xy}$ (right) at $T = 1.5$K as a function of $V_g$ and $V_b$ at $B = 2$T (**a**), 3T (**b**), 6T (**c**) and 8T (**d**) (device 2).

Dashed lines mark the phase boundaries corresponding to those in Fig. **1f**. The trion regions disappear above 6T due to the spin-valley Zeeman effect. Note that $R_{xy}$ is negligible inside the excitonic insulator region at high field due to absence of free carriers. Quantum oscillations in the W-layer are more pronounced at 8T and the excitonic insulator extends further into the pn region.

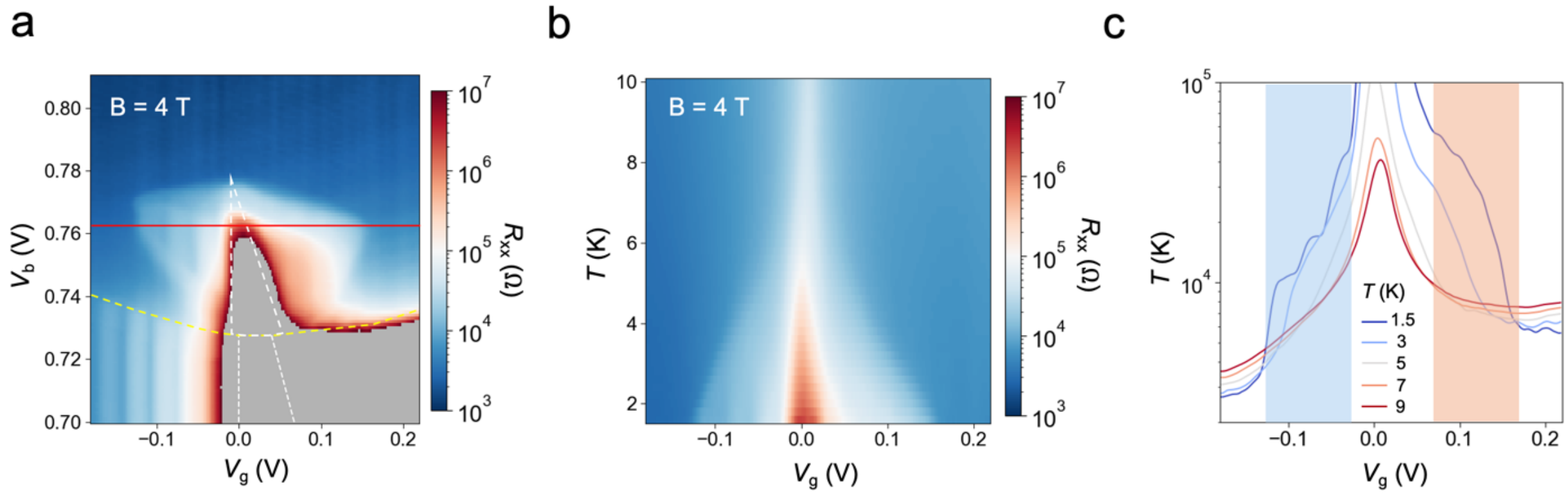

**Extended Data Figure 7 | Temperature dependence of $R_{xx}$. a,** $R_{xx}$ at $T = 1.5$K as a function of $V_g$ and $V_b$ under $B = 4$T (device 1). **b,** $R_{xx}$ as a function of $V_g$ and $T$ along the red line in **a**. Insulating behaviors around $V_g = 0$V emerge below $T \approx 8$K implying trion formation. **c,** Line cuts of $R_{xx}$ versus $V_g$ from **b** at varying temperatures. The blue- and the orange-shaded regions mark approximately the positive and negative trion regions, respectively.

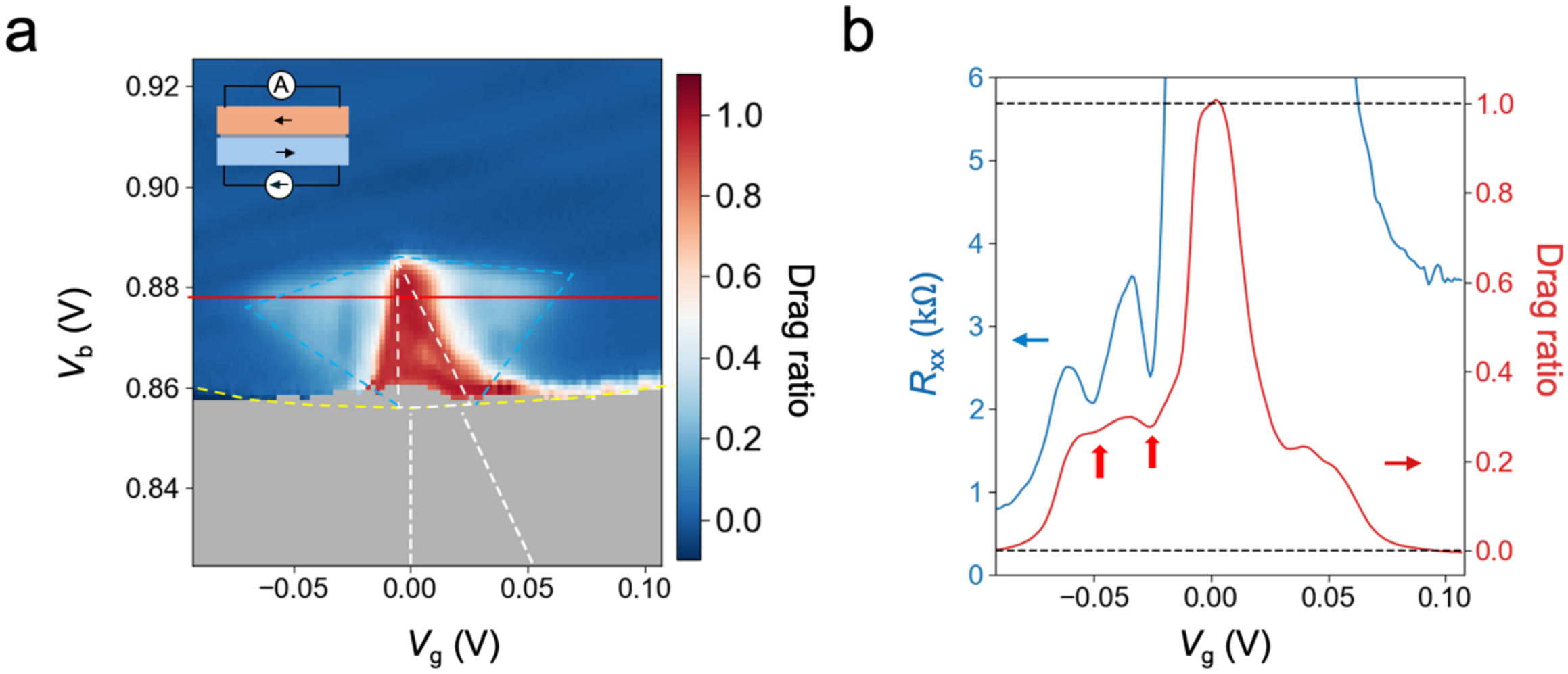

**Extended Data Figure 8 | Drag ratio and $R_{xx}$. a,** Drag ratio (drag current divided by the drive current) as a function of $V_b$ and $V_g$ at $T = 1.5$K and $B = 4$T (device 2). The inset shows the measurement configuration. The excitonic insulator region exhibits perfect drag and the trion regions also show strong Coulomb drags. Quantum oscillations are also observed inside the trion regions. **b,** Line cut of drag ratio and $R_{xx}$ along the horizontal red line in **a**. The dips in $R_{xx}$ and drag ratio (marked by red arrows) coincide with each other.